\documentclass[11pt,twoside]{article}

\usepackage{asp2006}
\usepackage{epsf}
\usepackage{psfig}
\usepackage{lscape}
\usepackage{graphicx}

\newcommand{\NIR}{{\sc Nir}}

\newcommand{\kms}{km\,s$^{-1}$}

\newcommand{\Ha} {H$\alpha$}

\newcommand{\HI} {\textsc{H\,i}}
\newcommand{\HII} {\textsc{H\,ii}}

\begin{document}
\title{The environment of nearby Blue Compact Dwarf Galaxies}   %%% Fill in title

\author{\'Angel R. L\'opez-S\'anchez\altaffilmark{1}, B\"arbel Koribalski\altaffilmark{1}, Janine van Eymeren\altaffilmark{2},  C\'esar %%@
Esteban\altaffilmark{3}, Attila Popping\altaffilmark{1,4},  \& John Hibbard\altaffilmark{5}}

\altaffiltext{1}{CSIRO / Australia Telescope National Facility, Sydney, Australia}
\altaffiltext{2}{Jodrell Bank Centre for Astrophysics, School of Physics \& Astronomy, Un. of Manchester, UK}
\altaffiltext{3}{Instituto de Astrof\'{\i}sica de Canarias, Tenerife, Spain}
\altaffiltext{4}{Kapteyn Astronomical Institute, University of Groningen, the Netherlands}
\altaffiltext{5}{National Radio Astronomy Observatory, USA}

%\ead{\mailto{Angel.Lopez-Sanchez@csiro.au}}

\begin{abstract} %%% Abstract to run on from here.

%Blue compact dwarf galaxies (BCDGs) represent the subset of 
%low-luminosity galaxies undergoing a strong and short-lived episode 
%of star formation at the present time. 
We are obtaining deep multiwavelength data of a sample of nearby blue compact dwarf galaxies (BCDGs) combining broad-band optical/NIR and H$\alpha$ %%@
photometry, optical spectroscopy and 21-cm radio observations. 
%in order to understand their chemical and physical properties, star formation 
%activity, kinematics, estimate the importance of the young/old 
%stellar populations within them and the environment in which they reside. 
%The selected BCDGs are found in different environments, 
%from apparently isolated to compact groups. 
Here we present \HI\ results obtained with the \emph{Australia Telescope Compact Array} for some BCDGs, all showing evident interaction features in %%@
their neutral gas component despite the environment in which they reside.
Our analysis strongly suggests that interactions with or between low-luminosity dwarf galaxies or HI clouds are the main trigger mechanism of the %%@
star-forming bursts in BCDGs; however these dwarf objects are only detected when deep optical images and complementary HI observations are performed. %%@
Are therefore BCDGs real isolated systems?

\end{abstract}

%%% MAIN BODY OF TEXT GOES HERE. CONSULT "INSTRUCTIONS FOR AUTHORS USING
%%% LATEX2E MARKUP", SECTIONS 2.3-2.6 FOR HELP WITH EQUATIONS, FIGURES,
%%% AND TABLES.

%\section{}   %%% Top level section head (remove "%" symbol)
%\subsection{}   %%% Second level section head (remove "%" symbol)
%\subsubsection{}   %%% Lowest level section head (remove "%" symbol)
%\section*{}    %%% Unnumbered top level section head (remove "%" symbol)
%\subsection*{}   %%% Unnumbered second level section head (remove "%" symbol)

\section{Introduction\label{S1}}

Blue compact dwarf galaxies (BCDGs) are a subset of low-luminosity galaxies undergoing strong and short-lived episodes of star formation at the %%@
present time. They usually exhibit a compact, irregular morphology and display intense, narrow emission lines superposed on a blue continuum.
Due to the low metallicity 
%($\sim$10 \% solar) 
and the considerable gas consumption involved in the violent star-formation, it is mostly believed that unlike spirals the star formation in BCDGs %%@
takes place in transient, sporadic bursts. 
The origin and nature of their starburts is still, however, poorly understood. There is increasing observational evidence that interactions of BCDGs %%@
with dwarf galaxies trigger the star formation activity in these systems [i.e. \cite{ME00};\cite{LSE08};\cite{LSE09}]
Optical images of BCDGs rarely show clear interaction features,
%Furthermore, hierarchical formation models of galaxies 
%\cite{KW93};\cite{Springel05} 
%predict that interactions between dwarf galaxies are more 
%common at high redshifts. 
%Indeed, BCDGs were an order-of-magnitude more populous than 
%normal bright galaxies at z=0.3 to 0.5 \cite{BR92}. 
%Detailed analysis of local BCDGs would have an important impact 
%on our knowledge of the evolution of the galaxies.
%Although there has been an increasing amount of BCDG data in the last years, 
but interferometric \HI\ studies have resulted in surprises.
%which, 
%together with the molecular gas component, is the raw material 
%for the star formation. Furthermore, 
Indeed, neutral hydrogen gas is the best tracer for galaxy-galaxy interactions because it is more easily disrupted by tidal forces than the stellar %%@
disk. 
\HI\ observations also provide an estimate of 
the total dynamical mass and the neutral gas content. Combining \HI\ data with parameters such as the absolute luminosity, star formation rate, %%@
stellar and dust content or the oxygen abundance can furnish powerful clues about the nature and evolution of BCDGs.

\begin{figure}[t!]
\centering
\includegraphics[angle=0,width=\linewidth]{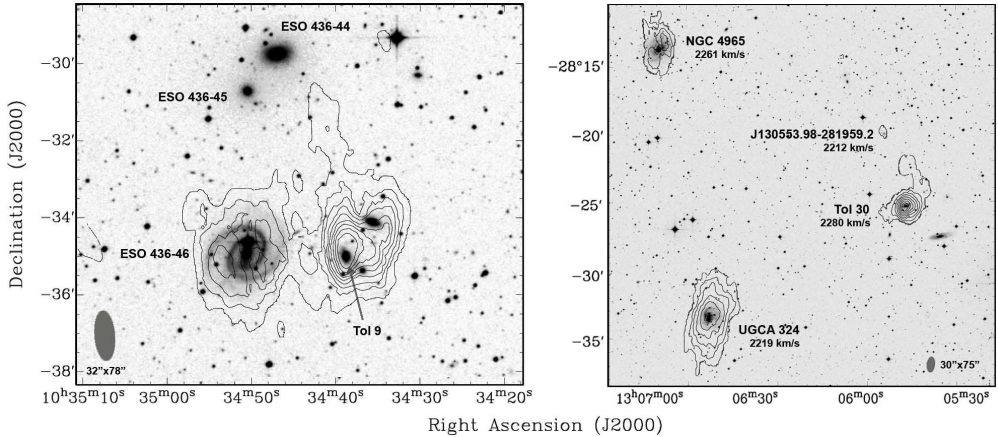}
\caption{\footnotesize{ATCA \HI\ distribution of the galaxy groups associated with Tol~9 [\emph{left}, \cite{LS08b}] and Tol~30 (\emph{right}) %%@
overlaid onto an $R$-band image from the DSS. }}
%Note that the angular resolution in both images is very similar but 
%we show a different area (the primary beam, the size of the 
%full field of view, is 33').}}
\label{fig1}
\end{figure}

\section{ATCA observations\label{S2}}

We obtained deep multi-wavelength data of a sample of nearby BCDGs combining broad-band optical/\NIR\ and \Ha\ photometry and optical spectroscopy %%@
with deep 21-cm radio continuum and line observations. The latter were obtained with the \emph{Australia Telescope Compact Array} (ATCA) %%@
interferometer using full synthesis observations (12 hours) in four different array configurations.
%consisting of the six antennae provides two different frequency bands. 
We selected 10 bright BCDGs that were detected in the \emph{\HI\ Parkes All-Sky Survey}, HIPASS; \cite{Barnes01};\cite{Koribalski2004}. Some BCDGs %%@
are apparently isolated (He 2-10, IC~4662, IC~4870, ESO~108-G017,  POX~4), while others belong to a galaxy pair
(TOL~1924-416, NGC~1510) or a galaxy group (Tol~9, Tol~30 and NGC 5253). 
%or a galaxy group. 
The radio data of the galaxies NGC~1510, NGC~5253 and IC~4662 are part of  \emph{The Local Volume \HI\ Survey} (LVHIS\footnote{LVHIS is a large %%@
project \cite{Koribalski08} that aims to provide detailed \HI\ distributions, velocity fields and star formation rates for a complete sample of %%@
nearby, gas-rich galaxies belonging to the Local Volume ($\sim$10 Mpc).
%(LV), the sphere of radius 10 Mpc centred on the Local Group. 
%With the ATCA we observed all LV galaxies that were detected in 
%HIPASS and reside south of approx. --30$^{\circ}$ declination. 
See \emph{http://www.atnf.csiro.au/research/LVHIS} for details.}). 
%We are using full synthesis observations (12 hours) in four different array 
%configurations of the ATCA interferometer to get deep \HI\ data. 
The final radio maps will be obtained by combining all available arrays, having both high sensitivity (\HI\ column density of $\sim5\times10^{19}$ %%@
cm$^{-2}$ for a 40" beam) and good angular ($\sim20-30$") and velocity ($10-20$ km\,s$^{-1}$) resolution. 
%We can achieve a range of angular resolutions by weighting the 
%data differently, thereby affecting the image sensitivity.  

\section{BCDGs in galaxy groups and galaxy pairs\label{S3}}

One of the key points in our analysis is to compare the properties derived for BCDGs belonging to galaxy groups with those that are in pairs or %%@
apparently isolated.  
Tol 9 is located in the center of the Klemola 13 group, its optical properties were published in \cite{LSE08};\cite{LSE09}. The \HI\ map %%@
(Figure~\ref{fig1}, \emph{left}) shows that the neutral gas is found in two regions: one associated with the spiral galaxy ESO 436-46 and the other %%@
embedding Tol 9 and two nearby objects, revealing a long \HI\ tail in direction to ESO 436-44 and ESO 436-45, galaxies mainly composed by old stellar %%@
populations. 
%The \HI\ kinematics are also very intriguing, because 
The \HI\ cloud in which Tol 9 an its surrounding dwarf galaxies are embedded seems to rotate as a single entity \cite{LS08b}. 
The kinematics of the tail suggest that it is of tidal origin.

Our analysis of the \HI\ gas in Tol 30 has revealed two faint tails starting at opposite sides of the galaxy, where the brightest \HII\ regions are %%@
located (Figure~\ref{fig1}, \emph{right}). The northern tail hosts around 15\% of the total \HI\ mass of the system. We detected a detached \HI\ %%@
cloud at the NE of Tol 30 that seems to show rotation. Our deep optical images confirm the detection of an object within this \HI\ cloud. We think it %%@
is not a tidal dwarf galaxy but an independent nearby low-luminosity dwarf galaxy that interacted with Tol~30.

The analysis of the \HI\ kinematics in NGC~5253 reveals a velocity gradient along the optical minor axis of the galaxy; it does not show any sign of %%@
regular rotation \cite{LS08a}.
Some authors suggested that this feature is an outflow. However, we think that its origin may be the disruption/accretion of a dwarf gas-rich %%@
companion or the interaction with another galaxy in the M\,83 subgroup. 
%Our finding of a distorted \HI\ morphology in the external 
%parts of the galaxy supports this hypothesis. 
%This fact and the finding of a kind of distorted \HI\ morphology in 
%the external parts of the galaxy may suggest an interaction 
%with the nearby spiral M\,83 in the past.

\begin{figure}[t!]
\centering
\includegraphics[angle=0,width=\linewidth]{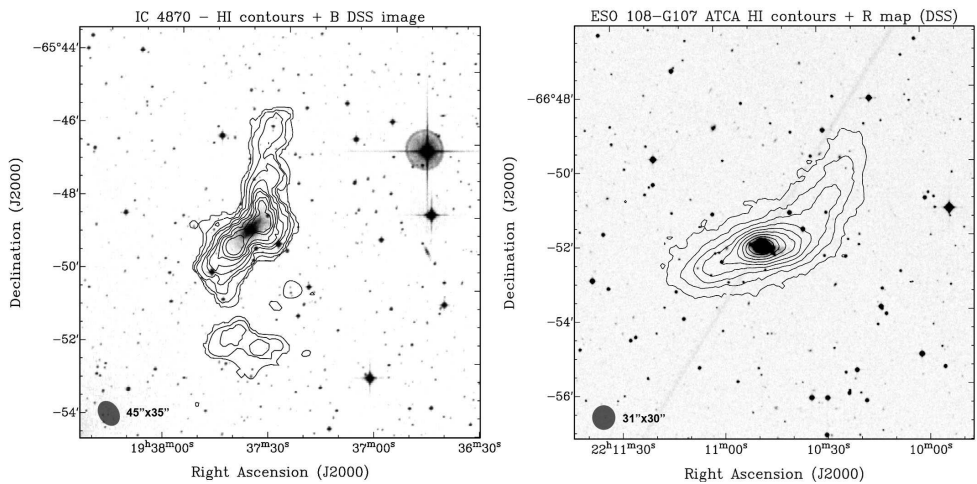}
\caption{\footnotesize{Preliminary ATCA \HI\ maps of the galaxies IC 4870 (\emph{left}) and ESO 108-G107 (\emph{right}) overlaid onto optical DSS %%@
images. }}
%Both \HI\ distributions show interaction features: IC 4870 seems to 
%be a merger, and ESO 108-G107 possesses a prominent \HI\ 
%tail with a peculiar kinematics.}}
\label{fig2}
\end{figure}

%\section{BCDGs in galaxy pairs\label{S4}}

Two of the BCDGs in our sample reside in a galaxy pair. The \HI\ map of TOL~1924-416 reveals a huge amount of neutral gas and an \HI\ bridge towards %%@
the companion galaxy \mbox{ESO 338-004B}. The \HI\ bridge has $\sim$33\% of all the neutral gas of the galaxy pair, 
and it has probably expelled from TOL~1924-416. A detailed analysis of the neutral gas of the galaxy pair NGC~1512 and the BCDG NGC~1510 is presented %%@
in \cite{KLS09}. Our data strongly support the interaction-induced scenario to explain both the \HI\ features and the star-formation activity in the %%@
system, that seems to be experiencing the first stages of a minor merger.

\section{Apparently isolated BCDGs\label{S5}}

The optical appearance of IC~4870 shows a compact star-forming core %(35") 
embedded in an elliptical low-luminosity component 
%with a size of $1.4'\times0.4'$. However, 
but its \HI\ map (Figure~\ref{fig2}, \emph{left}) reveals a lot of neutral gas and two long tails 
%with sizes of $\sim3.7'$ (northern tail) and $\sim4.2'$ (southern tail) 
arising in opposite directions. We detect two maxima of the \HI\ emission located at the beginning of the tails. 
The northern tail is quite straight and does not show important variations 
in its kinematics, but the southern tail is slightly curved towards the W. 
%The southern 
This tail possesses a knot at the south that hosts $\sim$14\% of all the \HI\ mass of the system 
%(it may seem detached from the tail if the \HI\ image is not deep enough) 
and shows a velocity gradient. All these facts suggest that IC\,4870 is experiencing a merging of two independent \HI\ clouds, being the origin of %%@
its strong star-formation activity.    

The \HI\ distribution found in the BCDG ESO 108-G107 (Figure~\ref{fig2}, \emph{right}) is more than 5 times the optical size. We detected a long tail %%@
towards the NW that has peculiar kinematics and is not aligned with a faint optical tail found at the W of the BCDG. The \HI\ distribution and %%@
kinematics of the eastern neutral gas also suggest the presence of a tail in this area.

The analysis of the \HI\ properties of the BCDG IC 4662 is presented in \cite{vEKLS09}.
%\cite{vEKLS09}. 
%The neutral gas distribution consists of two parts: the inner high 
%column density system (that coincides with its optical extent) is 
%perpendicular to the outer low column density system ending in a 
The low column density gas shows a
kind of tail towards the east. 
The kinematics are very disturbed: the overall velocity gradient runs from the north-east with velocities of 220 \kms\ to the south-west with %%@
velocities of 380 \kms, changing direction by about 90$^{\circ}$  in the centre of IC 4662. 
%The neutral gas distribution shows a kind of tail towards the E, 
%being the \HI\ kinematics of all the system very disturbed. 
%The overall \HI\ velocity gradient runs from the north-east with 
%velocities of 220 \kms\ to the south-west with velocities of 380 \kms. 
%It seems as if the
%direction of the rotation changes from the western part of
%the galaxy to the eastern part, causing a sudden change of
%the position angle of almost 90..
The chemical properties of some of the \HII\ regions are also intriguing and may suggest the presence of two objects that have experienced different %%@
chemical evolution.
%- You should cite the Hidalgo-Gamez et al. (2001) (A&A, 376, 386) 
%paper when talking about the chemical properties of IC 4662.

\section{Conclusions\label{S6}}

We performed a multi-wavelength analysis of some nearby BCDGs combining broad-band optical/NIR and H$\alpha$ photometry, optical spectroscopy and %%@
21-cm radio observations. 
We show that the \HI\ data are fundamental to understand their nature and dynamical evolution. All analyzed BCDGs show interaction features despite %%@
if they are located in a galaxy group, a galaxy pair or are apparently isolated. Our study confirms  \cite{LSE08} and \cite{LSE09} conclusions that %%@
interactions with or between low-luminosity dwarf galaxies are the main trigger mechanism of starbursts, specially on BCDGs. However these dwarf %%@
objects are only detected when deep optical images and complementary \HI\ observations are obtained. Therefore, BCDGs seem not to be real isolated %%@
systems.

%\acknowledgements %%% Text of acknowledgements runs on after this command.

%%% THE BIBLIOGRAPHY
%%%
%%% CONSULT SECTION 3 OF "INSTRUCTIONS FOR AUTHORS" FOR HOW TO USE NATBIB.
%%% AUTHORS ARE ENCOURAGED TO USE EITHER THE "THEBIBLIOGRAPY" ENVIRONMENT
%%% BY UNCOMMENTING (DELETING THE "%" SYMBOL) THE COMMANDS BELOW, OR BY
%%% USING THE BIBTEX ENVIRONMENT. TO FIND OUT WHICH IS APPLICABLE TO YOUR
%%% CONTRIBUTION, CONSULT THE VOLUME EDITORS FOR YOUR PROCEEDINGS.
%%%

\end{document}